\documentclass[aps,prl,twocolumn,superscriptaddress]{revtex4}

\usepackage{graphicx}
\usepackage{dcolumn}
\usepackage{bm}
\usepackage{amsmath}

\begin{document}


\title{Relativistic Mean Field Theory of \\Surface Pion
Condensation in Finite Nuclei}


\author{Hiroshi Toki}
\email{toki@rcnp.osaka-u.ac.jp}
\affiliation{Research Center For Nuclear Physics (RCNP), Osaka University,
Ibaraki, Osaka 567-0047, Japan}
\author{Satoru Sugimoto}
\email{satoru@riken.go.jp}
\altaffiliation[on leave of absence from ]
{Research Center For Nuclear Physics (RCNP)}
\affiliation{Institute for Chemical and Physical Research (RIKEN), Wako,
Saitama 351-0198, Japan}
\author{Kiyomi Ikeda}
\email{k-ikeda@riken.go.jp}
\affiliation{Institute for Chemical and Physical Research (RIKEN), Wako,
Saitama 351-0198, Japan}


\date{\today}

\begin{abstract}
 We study the possible occurrence of surface pion condensation in
finite nuclei in the relativistic mean field (RMF) theory. We are
led to this conjecture due to the essential role of pions in
few-body systems and the recent (p,n) experiments performed at
RCNP for spin-isospin excitations of medium and heavy nuclei. We
calculate explicitly various N=Z closed shell nuclei with finite
pion mean field in the RMF framework and demonstrate the actual
occurrence of surface pion condensation.
\end{abstract}

\pacs{21.60.Jz, 21.30.Fe, 21.10.Dr}


\maketitle

The pion is conjectured by Yukawa as the mediator of
nucleon-nucleon interaction \cite{yukawa35}. The pion is
identified then as the Nambu-Goldstone mode when the underlying
symmetry, the chiral symmetry, is spontaneously broken
\cite{nambu60}.  The pion plays the central role in hadron
physics, particularly for the low energy phenomena, and now the
chiral perturbation approach with the pion as the essential degree
of freedom is the powerful tool for the study of hadron properties
and collisions.

On the other hand, since the establishment of the shell model, we
usually solve nuclear many-body problems in the model space, where
single particle states have good parities.  Because the parity of
a nucleon changes when it absorbs or emits a pion, we must include
the higher configurations like 2p-2h (2 particle--2 hole),
$\dotsc$, to incorporate the effect of the pion in the
parity-conserved space.  To avoid such complications we
renormalize the central and the spin-orbit interactions in nuclei
to incorporate the effect of the strong correlations caused by the
pion.  It means that we make a model space in which the parity is
conserved and then define the effective interaction acting between
nucleons in the model space. One of the main purpose of the
present study is to expand the model space for the pion to play
its important role explicitly and see the effect on nuclear
structure.

As for the importance of the pion, we remind ourselves of the
findings of few-body systems, where they are solved rigorously
without the restriction of model space with a realistic
nucleon-nucleon interaction. The calculations are performed in the
non-relativistic framework and hence the pion is hidden mainly in
the tensor force, which is much larger than the central force in
the one pion exchange interaction even up to the Compton length.
There are many extensive calculations using the variational
principle and sophisticated numerical techniques and the
calculated results compare with experiments very well
\cite{carlson98,suzuki98}.  The calculated results demonstrate a
dominant role of the tensor force in the few-body system.  Almost
a half of the attraction is caused by the tensor force, and hence
the pion exchange interaction \cite{akaishi86}. The recent
variational calculations of the Argonne group up to A=8 system
also demonstrate the dominant role of the pion and further the
importance of the three-body interaction, which also originates
from the pion \cite{pieper01}.

There are several experimental data demanding the important role
of pion in medium and heavy nuclei.  The (p,n) reactions on medium
and heavy nuclei demonstrate that only a half of the Gamow-Teller
($\sigma \tau$) strengths is carried by 1p-1h excitations, while
the rest is interpreted as carried by 2p-2h excitations due to the
coupling to 1p-1h states by the strong tensor force
\cite{wakasa97}. A further dramatic result is the ratio of the
longitudinal and the transverse spin responses being close to one,
while the strong pionic correlations ought to provide a large
enhancement in the longitudinal channel \cite{wakasa99}. In
addition, there appear large short range correlations, which bring
the single particle strengths up to highly excited states, seen in
the (e,e'p) experiments as caused by the strong tensor force.

All these facts suggest us to look into the possibility of finite
pion mean field in medium and heavy nuclear systems.  We have been
reluctant in doing this, since we have to break the parity and
isospin symmetries once the pion mean field becomes finite and
have to project out the states with these good quantum numbers in
order to get observables. There is also a common sense that pion
condensation does not happen in nuclear matter of the saturation
density \cite{toki79,oset82}. This fact does not mean, however,
that the pion mean field vanishes in finite nuclear system.  The
pion is a pseudoscalar meson; it couples with a nucleon through
the $\vec\sigma \cdot \vec\nabla$ coupling, and therefore the
source term of the pion field needs the parity mixing and the
density modulation \cite{takatsuka76}. In infinite matter, we have
to provide this density modulation for the pion to work, which
costs energy. In finite nuclear system, we have the nuclear
surface and automatically there is the necessary density
modulation for the pion to work. Hence, if the pion mean field
turns out finite in nuclear system, the pionic correlations
utilize the change of the density at the surface.  Hence, we would
like to name this phenomenon as surface pion condensation.

We start with writing the relativistic meson-nucleon Lagrangian
density, which naturally includes the pion term, $\pi$;
\begin{widetext}
\begin{eqnarray}
       {\cal L}& = &{\bar\psi} [i \gamma^{\mu}\partial_{\mu}
                  - M-g_{\pi} \gamma_5 \gamma^{\mu} \tau^a
                  \partial_{\mu} \pi^a -g_{\sigma}\sigma
                  - g_{\omega} \gamma^{\mu}
                  \omega_{\mu}- g_{\rho}\gamma_{\mu}\tau^{a}\rho^{a \mu}
                  -e \gamma_{\mu} \frac{(1-\tau_{3})}
                  {2} A^{\mu} ]
                  \psi\nonumber\\
                  &&+ \frac{1}{2}\,
                  \partial_{\mu}\pi^a\partial^{\mu}\pi^a-
                  \frac{1}{2}m_\pi^2\pi^a\pi^a
                  + \frac{1}{2}\, \partial_{\mu}\sigma\partial^{\mu}\sigma
                - \frac{1}{2}m_\sigma^2{\sigma}^{2} - \frac{1}{3}g_{2}\sigma
                  ^{3} - \frac{1}{4}g_{3}\sigma^{4} \nonumber\\
                &&-\frac{1}{4}H_{\mu \nu}H^{\mu \nu} + \frac{1}{2}m_{\omega}
                   ^{2}\omega_{\mu}\omega^{\mu} + \frac{1}{4} c_{3}
                  (\omega_{\mu} \omega^{\mu})^{2}
                  -\frac{1}{4}G_{\mu \nu}^{a}G^{a\mu \nu}
                  + \frac{1}{2}m_{\rho}
          ^{2}\rho_{\mu}^{a}\rho^{a\mu}-\frac{1}{4}F_{\mu \nu}F^{\mu \nu}
,
\label{eq: lag}
\end{eqnarray}
\end{widetext}
where the field tensors $H$, $G$ and $F$ for the vector fields
are defined through
\begin{subequations}
\begin{eqnarray}
                 H_{\mu \nu} &=& \partial_{\mu} \omega_{\nu} -
                       \partial_{\nu} \omega_{\mu}, 
               \\
                 G_{\mu \nu}^{a} &=& \partial_{\mu} \rho_{\nu}^{a} -
                       \partial_{\nu} \rho_{\mu}^{a}
                     -g_{\rho}\,\epsilon^{abc} \rho_{\mu}^{b}
                    \rho_{\nu}^{c}, 
                    \\
                  F_{\mu \nu} &=& \partial_{\mu} A_{\nu} -
                       \partial_{\nu} A_{\mu}.
\end{eqnarray}
\end{subequations}
The pion term couples with the nucleon through the pseudo-vector
coupling. We take here all the terms used in Ref.
\cite{sugahara94}. Here, $\sigma$ denotes the scalar meson,
$\omega$ the vector meson and $\rho$ the isovector-vector meson.
The photon is denoted by $A$.  At the same time, in principle,
there should be a coupling of pion with the delta state.  In this
paper, we do not include this contribution for simplicity of
discussion, although this contribution will act constructively for
surface pion condensation.  We do not consider the tensor coupling
term of the $\rho$ meson either in order to concentrate on the
pion degree of freedom.  All these effects will be worked out in
the near future.

We then assume that the expectation value of the pion field is
finite.  In addition, the nuclear system is isospin-singlet due to
the choice of N=Z nuclei and hence the self-consistent Hamiltonian
is invariant under the isospin rotation \cite{ripka68}.  We choose
therefore the finite value of the pion mean field as the
z-component; i.e. $a=0$, without the loss of generality . We write
the finite pion mean field as $\pi$ without the isospin suffix.

For simplicity, we write only the equations of motion for the
nucleon, the sigma and pion with only the linear terms. They are
written as,
\begin{equation}
[i \gamma^{\mu}\partial_{\mu}
                  - M-g_{\sigma}\sigma- g_{\pi} \vec \nabla  \pi
                  \gamma_5 \vec\gamma\tau_0
                  ]\psi = 0,
\label{eq: dirac}
\end{equation}
for nucleon and
\begin{equation}
(\vec \nabla^2 - m_{\pi}^2)\pi= - g_{\pi}\vec
\nabla\langle{\bar\psi}\gamma_5 \vec \gamma\tau_0\psi\rangle,
\label{eq: k-gpi}
\end{equation}
and
\begin{equation}
(\vec \nabla^2 - m_{\sigma}^2)\sigma=g_{\sigma}\langle{\bar\psi}
\psi\rangle, \label{eq: k-gsigma}
\end{equation}
for pion and sigma mesons.  Here, the bracket $\langle \cdots
\rangle$ above denotes the ground state expectation value. The
other mesons follow the same equations of motion as above.

The equations (\ref{eq: dirac}) and (\ref{eq: k-gpi}) tell the
structure of surface pion condensation.  The pion field is
generally finite when the source term breaks the parity. The pion
field is enhanced by a spacial change of the source term. When the
pion field is finite in (\ref{eq: dirac}), the nucleon single
particle state breaks the parity, which provides again the pion
source term finite. The self-consistency provides the converged
solution to the above equations.

These equations tell the reason why we have not included the pion
mean field until now. The violation of the parity is caused by the
pion term in the above Dirac equation for nucleons. Hence, the
single particle state can be expressed as
\begin{equation}
\psi_{njm}(x)=\sum_{\kappa}W^n_{\kappa}\phi_{\kappa jm}(x).
\label{eq: psi}
\end{equation}
Here, $\phi_{\kappa jm}$ denote nucleon single particle wave
functions with the total angular momentum $j$ and its projection
$m$. We assume the spherical symmetry for the intrinsic state. The
summation over $\kappa$ means the parity mixing, where $\kappa$ is
$\kappa=-(l_\uparrow+1)$ for $l_\uparrow=j-1/2$ and
$\kappa=l_\downarrow$ for $l_\downarrow=j+1/2$. The calculational
detail will be provided in the forthcoming publication
\cite{sugimoto00}.

\begingroup
\squeezetable
\begin{table*}
 \caption{The contributions of kinetic energy (KE), sigma
($V_\sigma$), omega ($V_\omega$) and pion ($V_\pi$) energies to
the total binding energy (BE) given in MeV. Many N=Z closed shell
nuclei are calculated with and without finite pion mean field.
$^{12}$C, $^{56}$Ni, $^{100}$Sn and $^{164}$Pb are the jj-closed
shell nuclei and $^{16}$O, $^{40}$Ca and $^{80}$Zr are the
LS-closed shell nuclei.\label{tab:etotpi}}
 \begin{ruledtabular}
  \begin{tabular}{crrrrrrrrr}
  &\multicolumn{5}{c}{With $\pi$ mean field}&
   \multicolumn{4}{c}{Without $\pi$ mean field}
  \\
   \cline{2-6} \cline{7-10}
  &
  \multicolumn{1}{c}{BE}&
  \multicolumn{1}{c}{KE}&
  \multicolumn{1}{c}{$V_\sigma$}&
  \multicolumn{1}{c}{$V_\omega$}&
  \multicolumn{1}{c}{$V_{\pi}$}&
  \multicolumn{1}{c}{BE}&
  \multicolumn{1}{c}{KE}&
  \multicolumn{1}{c}{$V_\sigma$}&
  \multicolumn{1}{c}{$V_\omega$}\\
   $^{12}$C &   116 &    228 &  -1387 &   1196 &   -108 &
   98 &    197 &  -1376 &   1128 \\
   $^{16}$O &   148 &    227 &  -1794 &   1490 &    -14 &
   148 &    223 &  -1811 &   1498 \\
   $^{40}$Ca &   431 &    576 &  -4914 &   4127 &    -81 &
   427 &    556 &  -5003 &   4164 \\
   $^{56}$Ni &   650 &    955 &  -7431 &   6316 &   -281 &
   626 &    877 &  -7622 &   6340 \\
   $^{80}$Zr &   931 &   1118 & -10458 &   8751 &    -43 &
   931 &   1106 & -10519 &   8784 \\
   $^{100}$Sn &  1231 &   1599 & -13742 &  11629 &   -326 &
   1214 &   1509 & -14006 &  11689 \\
   $^{164}$Pb &  2109 &   2520 & -22913 &  19352 &   -409 &
   2094 &   2409 & -23352 &  19533
  \end{tabular}
 \end{ruledtabular}
\end{table*}
\endgroup
We show the numerical results.  We take the TM1 parameter set of
Ref. \cite{sugahara94}.  As for the pion-nucleon coupling we take
the value of Bonn A potential \cite{brockmann90}, which
corresponds to taking $g_\pi=f_\pi/m_\pi$ and $f_\pi\sim 1$. We
stress again here that we use all the terms given in the
Lagrangian (\ref{eq: lag}). This means that the saturation
property is guaranteed and the bulk part of the nucleus tends to
have the saturation density.  Since we are especially interested
in the occurrence of finite pion mean field and want to see its
effect under the simplest condition, we neglect the Coulomb term.
We calculate the N=Z closed shell nuclei as $^{12}$C, $^{16}$O,
$^{40}$Ca, $^{56}$Ni, $^{80}$Zr, $^{100}$Sn and $^{164}$Pb.

We show the results in Table~\ref{tab:etotpi} with and without
pion mean field. We show the total binding energy (BE), the
kinetic energy (KE), the sigma energy (V$_\sigma$), omega energy
(V$_\omega$) and pion energy (V$_\pi$). The finite pion mean field
case provides larger total binding energy than the one without
pion mean field. The pion energy is attractive, while all the
other terms as the kinetic energy, the combined sigma and omega
energy do not favor finite pion mean field.

\begin{figure}
\includegraphics{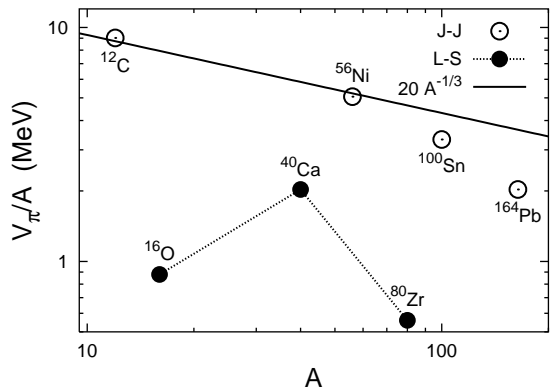}
\caption{\label{fig:pipar} The pion energy per nucleon as a function of
the mass number in the log-log plot.  There are two groups; one
is for the jj-closed shell nuclei denoted by open circle and the
other is for the LS-closed shell nuclei denoted by closed circle.
The pion energy per nucleon for the jj-closed shell nuclei
decreases monotonically and follows more steeply than $A^{-1/3}$,
which is shown by solid line.}
\end{figure}
We show the mass number dependence of the pion energy per nucleon
in Fig.~\ref{fig:pipar}. We mention here that the kinetic energy
and the sigma and omega energies are almost constant of the mass
number and hence they are volume like as easily recognized from
the numbers in Table~\ref{tab:etotpi}. We see, on the other hand,
a peculiar behavior in the pion energy. The magnitudes of the pion
energy are clearly separated into two groups.   One group is large
and the common feature is jj-closed shell nuclei; the magic number
nuclei due to a larger spin-orbit partner (j-upper) being filled.
The other group is small and they are LS-closed shell nuclei. The
pion energy per nucleon for jj-closed shell nuclei decreases
monotonically with the mass number. The rate of the decrease
follow more strongly than $A^{-1/3}$. This means that the pion
mean field energy behaves in proportion with the nuclear surface
or even stronger than that. Hence, we use the word of surface pion
condensation. Concerning the separation into two groups for the
pion energy; LS-closed and jj-closed shell cases, we shall mention
the possible reason later.

We discuss here the role of the pion by performing the parity
projection from the symmetry broken intrinsic state.  We write the
single particle state with mixed parity (\ref{eq: psi}) in a
simpler form as
\begin{equation}
|\bar {jm}\rangle = \alpha_j |{jm}\rangle+\beta_j |\tilde
{jm}\rangle .
\end{equation}
Here, $|\bar {jm}\rangle$ denotes a parity mixed single particle
state expressed as a linear combination of $|jm\rangle$, some
parity state (we call it as a normal parity state) and $|\tilde
{jm}\rangle$, the opposite parity state (abnormal parity state).
We write the intrinsic state with these single particle states up
to the Fermi surface and with all the magnetic sub-shells filled
as
\begin{align}
\Psi=&\prod_{jm}(\alpha_j |{jm}\rangle+\beta_j |\tilde
{jm}\rangle) \notag \\
=&\prod_{jm} \alpha_j |{jm}\rangle + \sum _{j_1m_1}\prod_{{jm}\neq
j_1m_1}
\alpha_j \beta_{j_1}|{jm}\rangle|\tilde {j_1m_1}\rangle
\label{eq: state}\\
+& \sum _{j_1m_1 j_2m_2}\prod_{{jm}\neq j_1m_1 j_2m_2} \alpha_j
\beta_{j_1}\beta_{j_2}|{jm}\rangle|\tilde {j_1m_1}\rangle|\tilde
{j_2m_2}\rangle+ \dotsb \notag
\end{align}
This intrinsic state has the total spin 0 because all the magnetic
sub-shells are filled, but the parity is mixed.  The first term,
$\prod_{jm} \alpha_j |{jm}\rangle$, in (\ref{eq: state}) has the
positive parity and corresponds to the ground state in the zeroth
order. The second term has the negative parity, since each normal
parity state, $|{j_1m_1}\rangle$, is replaced by an abnormal
parity state, $|\tilde {j_1m_1}\rangle$ for all occupied
$|{j_1m_1}\rangle$. Hence, if we say the first term as the 0p-0h
state, then the second term is a coherent 1p-1h state with $0^-$
spin parity.  The third term consists of 2p-2h states with a pair
of 1p-1h states with $0^-$ spin parity and therefore has $0^+$
spin parity.  The next term has three 1p-1h states with $0^-$ spin
parity and therefore has $0^-$ spin parity and so on.

Hence the positive parity projection $P_+$ would provides the
state with even number of 1p-1h states with $0^-$ spin parity.
$P_+ \Psi=|0\rangle+|2p-2h\rangle+|4p-4h\rangle+\dotsb$. This
means that the positive parity projection provides 2p-2h states as
the major correction terms. Hence, surface pion condensation
together with parity projection provides the 2p-2h admixture due
to the pion exchange interaction as the case of the $\alpha$
particle \cite{akaishi86}. The negative parity projection $P_-$
would provide the state with odd number of 1p-1h states with $0^-$
spin parity. $P_- \Psi=|1p-1h\rangle+|3p-3h\rangle+\dotsb$. This
is the brother state having the quantum number of $0^-$ to the
$0^+$ ground state.  The ground state consists of highly
correlated particle--hole states and the $0^-$ state is,
therefore, a coherent 1p-1h state compiled on the highly
correlated $0^+$ ground state.

This demonstration gives us the hint of providing the two groups
of the strong pion energy case, the jj-closed shell, and the weak
pion energy case, the LS-closed shell.  In the LS-closed shell
case, the higher spin-orbit partner in the upper shell above the
Fermi surface is not used for the $0^-$ particle-hole excitations.
On the other hand, in the jj-closed shell case, this higher
spin-orbit partner state is now filled and is used for the $0^-$
particle-hole excitations as a hole state.  Hence, more attraction
is expected for jj-closed shell nuclei than those of the LS-closed
shell nuclei.  The pion energy is gained largely for the jj-closed
shell nuclei, because the number of states to be mixed by the
pionic correlations increases and these states contribute
coherently to the pion energy.  We are, however, not yet clear
about the peculiar mass dependence of the LS-closed shell nuclei.
This distinct feature of the difference between the LS-closed and
the jj-closed shell nuclei can be the consequence of using only
the pion correlations.

We discuss here the qualitative consequence of surface pion
condensation. First, we discuss the Gamow-Teller (GT) transitions.
Without pion condensation, there exist no transitions for
LS-closed shell nuclei as $^{16}$O, $^{40}$Ca and only two
transitions, for example, for $^{90}$Zr. However, the mixing of
parity in the intrinsic state allows transitions of 2p--2h states.
This makes the spectrum of the GT transitions with some GT
strengths above the two dominant peaks in $^{90}$Zr
\cite{wakasa97}. Hence, naturally we have large strengths in the
continuum in the simple mean field theory as the experiment
demands.

The longitudinal spin response functions should be largely
modified due to pion condensation.  The longitudinal spin response
is caused by the pionic correlations.  Since the large pionic
strength is used up to construct the nuclear ground state, the
pionic fluctuation oughts to be reduced largely.  This should make
the spin response in the pion channel weak.  This remains to be
demonstrated in the future work.

The surface pion condensation provides us with the possibility to
describe the short range correlation effect. As seen above in the
discussion of the parity projection, the surface pion condensation
provides a large amount of the 2p--2h excitations in the nuclear
ground state automatically.  There should be many other
consequences of surface pion condensation in nuclear phenomena.
The pairing correlations and the spin-orbit couplings are all
surface phenomena and surface pion condensation would couple with
these correlations and would provide rich phenomena.

We have discussed the possible occurrence of surface pion
condensation in order to understand the recent (p,n) experimental
data taken at RCNP.  This suggestion is motivated by the missing
pion contribution in the discussion of ground states of finite
nuclei, while pions are essential for hadron physics.  We have
made calculations with the inclusion of finite pion mean field and
demonstrated the finite results for N=Z closed shell nuclei.  We
have demonstrated that the pion energy behaves proportional to or
even stronger than the nuclear surface.  Hence the name surface
pion condensation is used in this paper.  We have made qualitative
discussions on the consequence of surface pion condensation on the
Gamow-Teller strengths, the spin response functions and the short
range correlations.  The large difference of the pion energies in
jj-closed shell and LS-closed shell nuclei has been found and may
be connected with the mechanism of a part of spin-orbit splitting
as due to the tensor force discussed long time ago by Takagi {\it
et al.} and Terasawa \cite{terasawa60}.  To go further, we have to
fix the parameters of the Lagrangian (\ref{eq: lag}), which
includes the effect of finite pion mean field and other terms as
the delta state and the $\rho$ meson tensor term.

\begin{acknowledgments}
We acknowledge fruitful discussions with H.~Horiuchi and
Y.~Akaishi on the roll of the pion in light nuclei.
\end{acknowledgments}


\end{document}